\documentclass[11pt]{evolution}    

\usepackage{natbib}

\usepackage{epsfig,amsmath,amsfonts,amssymb}

\newcommand{\be}{\begin{equation}}
\newcommand{\ee}{\end{equation}}
\newcommand{\bea}{\begin{eqnarray}}
\newcommand{\eea}{\end{eqnarray}}
\newcommand{\bml}{\begin{mathletters}}
\newcommand{\eml}{\end{mathletters}}

\newcommand{\no}{\nonumber}

\renewcommand{\citep}[1]{(\citealt{#1})}

\raggedright

\begin{document}

\title{Evolutionary advantage of small populations on complex fitness landscapes}
\author{Kavita Jain$^{1}$%
       \email{Kavita Jain - jain@jncasr.ac.in}%
      \and
         Joachim Krug$^2$%
         \email{Joachim Krug - krug@thp.uni-koeln.de}
       and 
         Su-Chan Park\correspondingauthor$^{2,3}$%
         \email{Su-Chan Park\correspondingauthor - spark0@catholic.ac.kr}%
      }

\address{%
    \iid(1)Theoretical Sciences Unit and  Evolutionary and
Organismal Biology Unit,
Jawaharlal Nehru Centre for Advanced Scientific Research, Jakkur P.O.,
Bangalore 560064, India\\
    \iid(2)Institut
f\"ur Theoretische Physik, Universit\"at zu K\"oln,
Z\"ulpicherstr. 77, 50937 K\"oln, Germany\\
    \iid(3)Department of Physics,
The Catholic University of Korea, Bucheon 420-743, Korea
}%

\maketitle
Running Title : Advantage of small populations
%Changed the running title

\vspace{5mm}
{\bf Contact Information (for all authors)}

Kavita Jain 

{\bf postal address}: Theoretical Sciences Unit and  Evolutionary and
Organismal Biology Unit,
Jawaharlal Nehru Centre for Advanced Scientific Research, Jakkur P.O.,
Bangalore 560064, India

{\bf work telephone number}: +91-80-22082948

{\bf E-mail}:jain@jncasr.ac.in

\vspace{5mm}
Joachim Krug  

{\bf postal address}: Institute f\"ur Theoretische Physik, Universit\"at zu 
K\"oln, Z\"ulpicher Str. 77, 50937 K\"oln, Germany

{\bf work telephone number}: +49-221-470-2818

{\bf E-mail}:krug@thp.uni-koeln.de

\vspace{5mm}
Su-Chan Park (Corresponding Author)

{\bf address}: Department of Physics, The Catholic University of Korea,
43-1 Yeokgok 2-dong, Wonmi-gu, Bucheon 420-743, Republic of Korea

{\bf work telephone number}: +82-2-2164-4524

{\bf E-mail}: spark0@catholic.ac.kr

\clearpage
\begin{abstract}
Recent experimental and theoretical 
studies have shown that small asexual populations evolving on complex
fitness landscapes may achieve a
higher fitness than large ones due to the increased heterogeneity of 
adaptive trajectories. Here we introduce a class of haploid
three-locus fitness landscapes that allow the investigation of this
scenario in a precise and quantitative way. 
Our main result derived analytically
shows how the probability of choosing the path of the 
largest initial fitness increase grows with the population
size. This makes large populations more likely to get trapped
at local fitness peaks and implies an advantage of small
populations at intermediate time scales. The range of
population sizes where this effect is operative coincides with
the onset of clonal interference. Additional studies
using ensembles of random fitness landscapes show that the
results achieved for a particular choice of three-locus landscape
parameters are robust and also persist as the number of loci increases.
Our study indicates that an
advantage for small populations is likely whenever the fitness landscape
contains local maxima. The advantage appears at intermediate
time scales, which are long enough for trapping at local
fitness maxima to have occurred but too short for peak escape by
the creation of multiple mutants.

\vspace{1cm}
\end{abstract}

\clearpage

Consider a population  experiencing a recent environmental change.
Assuming that the population is ill-adapted to the new
environment, as is typically the case in the beginning of an evolution 
experiment \citep{LT1994}, the adaptation to the new environment 
relies on the supply of beneficial mutations available to the
population. During the early stages of adaptation, it is a good
approximation to assume that the supply of beneficial mutations is
unlimited. Then as the rate at which the beneficial mutations appear
is given by the mutation rate per individual times the population
size, a large population is expected to experience more beneficial
mutations and hence adapt faster than a small population.  

\parindent=2.0em
This conclusion does not seem to depend on the topography and epistatic interactions in the fitness landscape. 
If the fitness landscape is non-epistatic in the sense that
(beneficial) mutations act multiplicatively on fitness, a large
population adapts faster than a small population, although this
advantage is strongly reduced in asexuals due to clonal interference
\citep{GL1998,deVisser1999,W2004,Park2007,PSK2010}. Even if the
fitness landscape is highly epistatic such as a maximally rugged
(\textit{house-of-cards})  landscape \citep{Kingman1978}, a large
population still wins on average \citep{Park2008}.  
This leaves the question as to whether small populations might be at
an advantage in complex fitness landscapes with an intermediate degree of
epistasis and ruggedness. 

In a recent experimental
work~\citep{Rozen2008} studying the adaptation dynamics  
of populations of \textit{E. coli} in simple and complex nutrient environments,
it was found  that small populations could attain higher fitness
than large populations in a complex medium which can be expected, on general grounds,
to give rise to a rugged and epistatic fitness landscape. 
The observed fitness advantage of small populations was associated with a greater heterogeneity in their 
adaptive trajectories compared to large populations. Specifically, 
small populations that eventually reached the highest fitness levels 
were often the ones that initially displayed a rather shallow fitness increase, whereas
the fitness of those that gained a large initial advantage tended to level off quickly.
In large populations trajectories were more uniform and typically showed a rapid
initial fitness increase followed by a significant slowing down or saturation. 

These experiments suggest that an adaptive advantage can arise from the 
higher level of stochasticity in the incorporation of beneficial mutations
displayed by small populations, provided the
topography of the underlying fitness landscape is sufficiently complex. 
As the detailed structure of the experimental fitness
landscape is unknown and unfeasible to determine, it is useful to
investigate this mechanism theoretically. In previous work this was done using 
extensive simulations for a class of random fitness landscapes with
tunable ruggedness \citep{Handel2009}. The main conclusion from this study
was that an advantage of small populations can be observed in a substantial fraction
of random landscapes, but the dependence of the effect on parameters such as 
the population size, the mutation rate and the fitness effect of beneficial mutations
was not explored systematically. 

In this article, we introduce a minimal, analytically tractable model which 
captures the dynamic behavior of the population fitness in the
experiments by \cite{Rozen2008}. 
We show that the simplest fitness landscape that can exhibit a small population
advantage is a haploid, diallelic three-locus landscape where the
genotypes of minimal and maximal fitness are separated by three
mutational steps. 
There are then 3! = 6 distinct shortest paths leading from
the global fitness minimum (the wild type) to the global maximum, corresponding to the
different orderings in which the mutations are introduced into the
population \citep{Gokhale2009}. We
distinguish between \textit{smooth paths} along which fitness
increases monotonically, and \textit{rugged paths} containing at least
one deleterious mutation. The three-locus landscape is constructed
in such a way that the rugged paths contain a local fitness maximum, and 
they confer the greatest initial fitness increase to a population initially
fixed for the minimal fitness genotype. 

The existence of rugged paths is the hallmark
of \textit{sign epistasis}, a specific type of genetic interaction
under which a given mutation can be beneficial or deleterious
depending on the genetic background \citep{Weinreich2005a,Poelwijk2007}. 
Sign epistasis implies that at least some of the mutational pathways 
leading to the global maximum of the fitness landscape are rugged,
and thus inaccessible to an adaptive dynamics that is constrained to increase
fitness in each step. In addition, sign epistasis is a necessary (but not sufficient) condition 
for the existence of multiple fitness maxima \citep{Poelwijk2011}. The presence of sign epistasis was 
established in several recent experimental studies, where all combinations of a selected
set of individually beneficial  or deleterious mutations
were constructed and their fitness effects (or some proxy thereof) 
were measured \citep{Weinreich2006,Lozovsky2009,dVPK2009,Carneiro2009}. 
There is also considerable evidence for the existence of multiple fitness maxima
from evolution experiments using bacterial and viral populations 
\citep{Korona1994,Burch1999,Burch2000,Elena2003}. It is therefore reasonable to assume that
sign epistasis and mulitple fitness maxima are present also 
in the complex environment considered by \cite{Rozen2008}.

The analysis presented below shows 
that the speed of adaptation is generally an increasing
function of population size both along smooth and along rugged paths. However, 
the \textit{probability} 
with which a particular type of path is chosen depends
on population size in such a way that small populations can be favored at least
over a certain range of time scales.  
In particular, as we shall show, the probability to choose
the rugged path in the three-locus model rises sharply with the onset of clonal interference,
and it approaches unity when the dynamics becomes 
completely deterministic for very large populations, because then
the mutation with the largest initial fitness increase is certain
to fix \citep{JK2007}. 
The dynamics of small populations are less predictable, and they
therefore enjoy an advantage by more frequently avoiding
getting trapped at the local fitness maximum. The main part of the
article is devoted to a detailed, quantitative analysis of this
scenario. We then show that the mechanism identified within the 
specific three-locus model is robust by simulating 
populations on variants of the house-of-cards fitness landscape with 
three or more loci, and conclude the paper with a discussion of our key results.

\section*{Models}

\subsection*{Fitness landscapes}

In the main part of this work 
we consider the space of genotypes composed of 
three loci with two alleles each, which will be denoted $0$ and $1$. 
Each genotype is assigned a fitness $W$ according to  
\begin{eqnarray}
W(000) &=& 1 \no \\
W(001)&=&W(010)=1+s_1 \no \\
W(100)&=&1+s_2 \no\\
W(011)&=&W(101)=W(110)=(1+s_1)^2 \no \\
W(111)&=&(1+s_1)^3
\label{model}
\end{eqnarray}
where $s_2 > s_1 > 0$ and $(1+s_1)^2 < 1+s_2 < (1+s_1)^3$.  
Thus there is a local fitness maximum at genotype $\{100 \}$ and the global maximum is
located at genotype $\{111\}$. 

In addition to the landscape equation (\ref{model}), we consider two
ensembles of random fitness landscapes
consisting of $L$-locus genotypes with two alleles at each site. In the first 
ensemble referred to as unconstrained ensemble, the 
least fit genotype is assigned the allele $0$ at every locus and
fitness $1$ while the rest of the 
genotypes are given fitnesses 
\begin{equation}
W(\textrm{genotype}) = 1 + S x,
\label{Eq:fit_HOC}
\end{equation}
where $S$ controls the strength of selection 
and $x$ is a random number drawn from an exponential distribution with
mean $1$. 
This is Kingman's house-of-cards model adapted to a finite number of diallelic loci
\citep{Kingman1978,Kauffman1987,JK2007,Park2008}. 
%Since the global maximum of the landscape~\eqref{model} 
%is located at the antipodal point, namely \{111\}, of the
%least fit genotype in the genotype space, i
It is also instructive to 
study a constrained version of the above model in which the fittest genotype 
has all loci with allele 1~\citep{Kloezer2008,Carneiro2009}. 
Such landscapes are generated by assigning the
maximum value amongst $2^L-1$ random numbers given by equation
\eqref{Eq:fit_HOC} as the 
fitness of the antipodal point  of the
least fit genotype in the genotype space. 
%as follows: At first, $2^{L}-1$random numbers, $1 + S x$, are drawn. The maximum value among these random numbers is assigned to the antipodal genotype and the other numbers are distributed to the genotypes randomly. 
Note that these landscapes have exactly the same statistics as those which are 
sampled under the condition that the antipodal genotype is the fittest
among all randomly generated landscapes by equation \eqref{Eq:fit_HOC}.

\subsection*{Population dynamics}

We mainly work with a finite
population of size $N$ which evolves according to standard Wright-Fisher
dynamics in discrete generations.  In each generation, an offspring
chooses a parent with a probability proportional to the parent's
fitness and copies the parent's genotype. Then the point mutation process is
implemented symmetrically in which $0 \leftrightarrow 1$ with  
probability $\mu$. This process is repeated until all $N$ offspring
have been generated. 
In the actual simulations, we treated the population size of each genotype
as a random variable which is sampled according to a multinomial
distribution; for details see \cite{Park2007}.

It is also useful to compare the results obtained for
finite $N$ with the predictions of the deterministic
mutation-selection dynamics of ``quasispecies'' type 
which applies for infinite populations
\citep{Buerger,JK2006}. This is done by
iterating the deterministic evolution equations for the 
frequencies $f(\sigma,t)$ of genotype $\sigma$ at
generation $t$, which read
\begin{equation}
\label{quasispecies}
f(\sigma,t+1)= \frac{\sum_{\sigma'} M(\sigma \leftarrow
  \sigma') W(\sigma')f(\sigma',t) }{\sum_{\sigma'} W(\sigma')f(\sigma',t)}.
\end{equation}
Here $M(\sigma \leftarrow  \sigma')= \mu^{d(\sigma,\sigma')}
(1-\mu)^{L-d(\sigma,\sigma')}$ is the probability that an $L$-locus 
genotype $\sigma'$ mutates to genotype $\sigma$ at Hamming distance
$d(\sigma,\sigma')$.

\section*{Results}

%--------------------------------------------------------------------------
\subsection*{Evolution time scales}
%\label{timescales}

We begin by a comparison of the time taken to reach the global maximum
along the smooth and the rugged paths on the fitness landscape defined by
equation (\ref{model}). 
A population that is initially fixed for the minimum
fitness genotype $\{000\}$ has a choice between three single site
mutations. Two of these (to genotypes $\{001\}$ and $\{010\}$) lead
the population to a smooth path towards the global fitness maximum 
$\{111\}$, whereas the third leads to the local fitness maximum
$\{100\}$ from which the population can escape only by the creation of
a double mutant \citep{Iwasa:2004,Weinreich2005b,Weissman2009}.

We estimate the time scales of the relevant 
evolutionary processes in the  strong selection weak mutation (SSWM)
regime \citep{G1983AmNat,Gillespie:1984,Orr2002}, 
where $N \mu \ll 1$ and the population is monomorphic almost all the time.
For simplicity, in the present discussion we neglect back mutations 
towards the wild type genotype $\{000\}$. Starting from the wild
type, each of the single step mutants
is generated in the population at rate $N \mu$ and goes to fixation
with probability $\pi(s)$ given by \citep{Kimura}
\be
\pi(s)\approx\frac{1-e^{-2s}}{1-e^{-2 N s}}
\label{WF}
\ee
with $s = s_1$ or $s_2$. For $N^{-1} \ll s \ll 1$, the fixation
probability 
%Throughout this
%subsection we will assume that $N^{-1} \ll s \ll 1$, so that
 $\pi(s) \approx 2s$. 
The waiting times for low fitness
($T_1$) and high fitness ($T_2$) mutants that will ultimately fix are
therefore  
\begin{equation}
\label{waitingtimes}
T_1 \approx \frac{1}{2 \mu N s_1}, \;\;\; T_2 \approx \frac{1}{2 \mu N s_2}.
\end{equation}
Adaptation along one of the smooth paths proceeds by sequentially fixing
two additional beneficial mutations with selection coefficient $s_1$,
and the total evolution time is therefore $T_\mathrm{smooth} \approx 3 T_1$. 

By contrast, populations that choose the rugged path need to escape
from the local fitness peak $\{100\}$ in order to reach the global
maximum. The corresponding escape time can be estimated 
along the lines of \cite{Weinreich2005b}.
Following these authors we introduce the selection coefficients
\begin{equation}
\label{sbensdel}
s_\mathrm{ben} = \frac{W(111)}{W(100)} - 1 \approx 3 s_1 - s_2, \;\;\;
s_\mathrm{del} = \frac{W(100)}{W(101)} - 1 \approx s_2 - 2 s_1
\end{equation}
which express the relative fitness advantage of the global maximum
compared to the local peak ($s_\mathrm{ben}$) and that of the valley
genotypes compared to the local peak ($s_\mathrm{del}$),
respectively. Depending on the population size, the peak escape can 
proceed through two distinct pathways. In populations smaller than a
critical size $N_c$ \citep{Weinreich2005b}, the two
mutations separating the genotypes $\{100\}$ and $\{111\}$ fix
sequentially, while in larger populations they fix simultaneously. We
are interested in population size $\gg N_c \approx \ln (s/\mu)/s$
which can be easily satisfied in the SSWM regime as $N s \gg 1$. 
In the simultaneous mode the escape time is given approximately by 
\begin{equation}
\label{Tesc}
T_\mathrm{esc} \approx \frac{s_\mathrm{del}}{4 N \mu^2 s_\mathrm{ben}}. 
\end{equation}
Assuming all selection coefficients $s_1, s_2, s_\mathrm{ben},
s_\mathrm{del}$ to be of a similar magnitude $s$, we see that
\begin{equation}
\label{timeratio} 
\frac{T_\mathrm{esc}}{T_{1,2}} \sim \frac{s}{\mu} \gg 1
\end{equation}
whenever $\mu \ll s$, 
which is expected to hold under most conditions. In particular,
it is true in the SSWM regime  because $N \mu \ll 1$ and $N s \gg 1$. 
Equation (\ref{timeratio}) implies
that the evolution time $T_\mathrm{rugged}$ along a rugged path is 
dominated by the escape time $T_\mathrm{esc}$, and is much larger 
than $T_\mathrm{smooth}$. However, both 
equation (\ref{waitingtimes}) and equation (\ref{Tesc}) 
share the same dependence on population size $N$, so once the type
of evolutionary path is chosen, a large population is always at a 
relative advantage.

%--------------------------------------------------------------------------
\subsection*{Mean fitness evolution and heterogeneous adaptive trajectories}

\begin{figure}[ht]
\includegraphics[width=\textwidth]{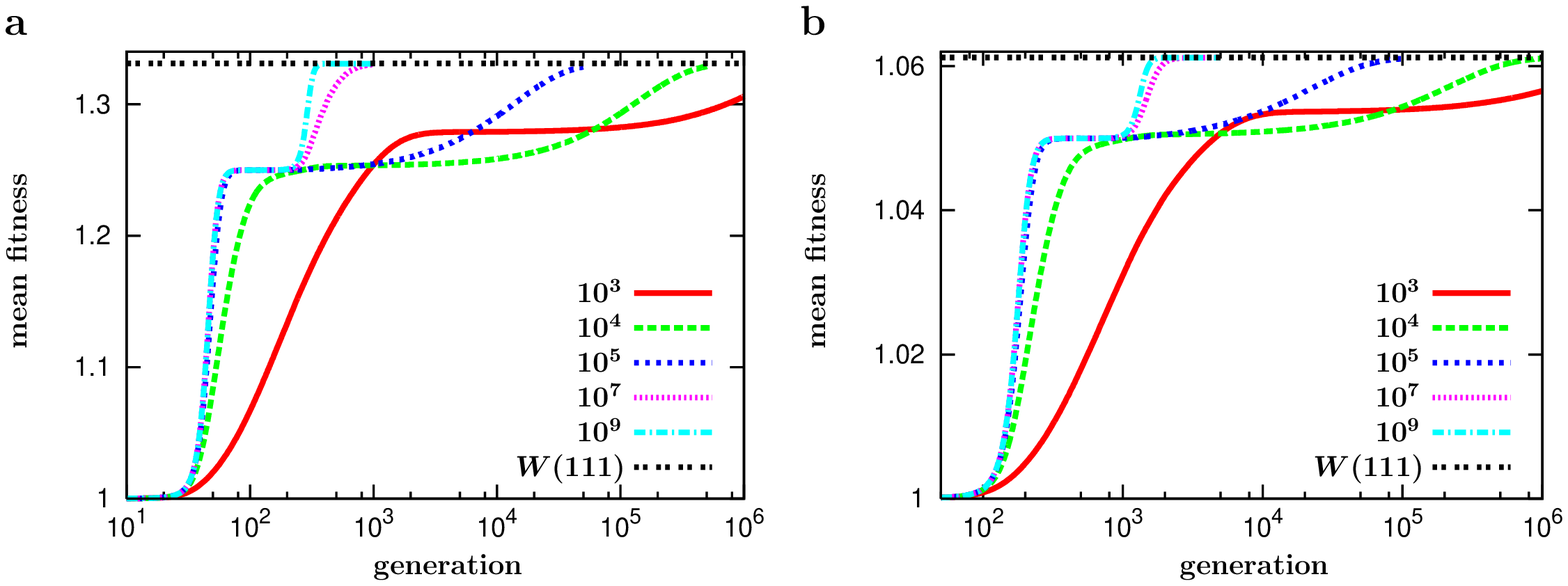}
\caption{\label{fit} Average fitness as a function of time on the
fitness landscape defined by equation (\ref{model}) for $\mu = 10^{-5}$,
(a) $s_1 = 0.1$, $s_2 = 0.25$ and (b) $s_1 = 0.02$, $s_2 = 0.05$, and
various population sizes indicated in the figures. 
As a guide to the eye, the maximum fitness values $W(111)$ 
for each case are also drawn.
The data have been averaged over $10^5$ histories. 
Fitness increases with population
size at short times and at long times, but in both cases this
relationship is reversed for a range of population sizes at intermediate times. 
}
\end{figure}

Figure~\ref{fit} shows the evolution of the population fitness
obtained from simulations of the Wright-Fisher model in the landscape
defined by equation (\ref{model}). Each curve contains data  
averaged over many stochastic histories for a given value of $N$, keeping other
parameters fixed, and starting with all individuals at the
genotype $\{000 \}$ with lowest fitness. 
At short times the fitness rises more rapidly for larger populations,
as expected on the basis of the estimates given in equation
(\ref{waitingtimes}) 
for $N \mu \ll 1$, while for $N \mu > 1$ fitness becomes independent of $N$.
Large populations are also seen to be at an advantage for extremely 
long times, beyond $10^5$ generations. However, for both parameter
sets displayed in the figure, the ordering of fitness with increasing
population size is reversed in an intermediate time interval, which
begins at around 2000 generations. 
 
\begin{figure}[ht]
\includegraphics[width=\textwidth]{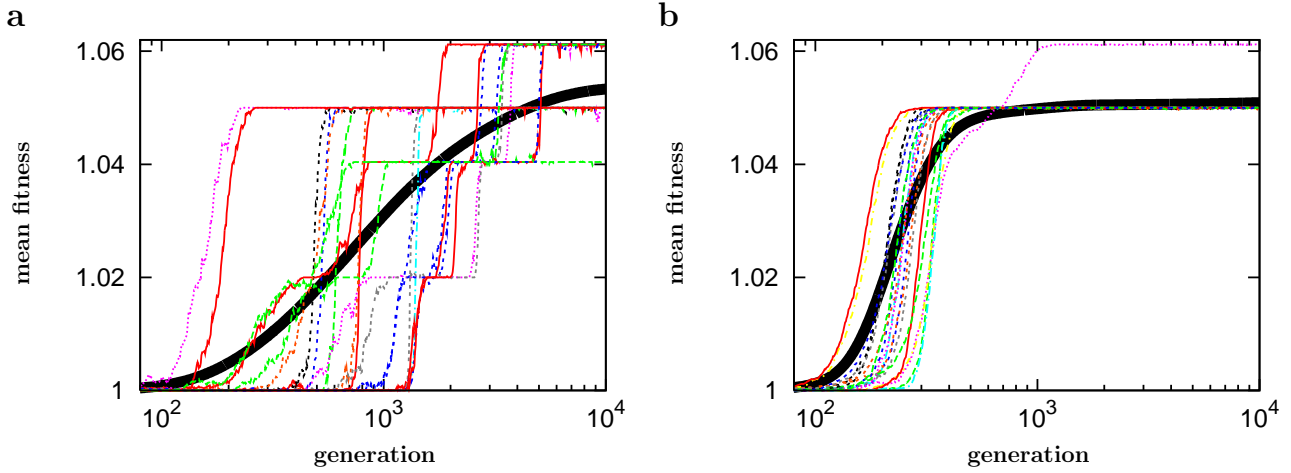}
\caption{\label{Whist}Population mean fitness as a function of time for 20 
histories on the fitness landscape equation (\ref{model}). 
The population size is (a) $N=10^3$ and (b) $10^4$, 
the selection coefficients are $s_1=0.02$ and $s_2=0.05$ and the mutation
probability is $\mu=10^{-5}$ (same as in Figure~\ref{fit}(b)). The smooth curves depict the average over
$10^5$ independent runs.}
\end{figure}

The origin of this reversal is illustrated in Figure~\ref{Whist}, 
which shows individual fitness trajectories for the parameter set
of Figure \ref{fit} (b) and two different population sizes. Individual
fitness trajectories display a step-like behavior, which reflect the 
transitions in the {\it most populated genotype}. In particular, populations 
in which the local peak genotype $\{100\}$ becomes dominant are seen
to remain trapped at the local peak for a long time (compare to 
equation (\ref{Tesc})). Although the initial rise in fitness is much faster 
for the large populations than for the small ones, the \textit{fraction}
of trajectories that take the rugged path (and thus get trapped at
the local peak) also increases with increasing $N$, from 11/20 in the 
Figure \ref{Whist}(a) ($N = 10^3$) to 19/20 in Figure \ref{Whist}(b)
($N=10^4$). As  
a consequence, the fitness after $10^4$ generations, when 
averaged over all trajectories, is larger for the small populations
than for the large ones.    
Similar to the experimental observations of \cite{Rozen2008} and the
simulations of \cite{Handel2009},  
smaller populations reach a higher fitness
level because their adaptive trajectories are more heterogeneous, 
allowing them to avoid trapping at the local fitness peak in a larger
number of trials.

\subsection*{\label{Pr}Path probability}

\begin{figure}
\includegraphics[width=\textwidth]{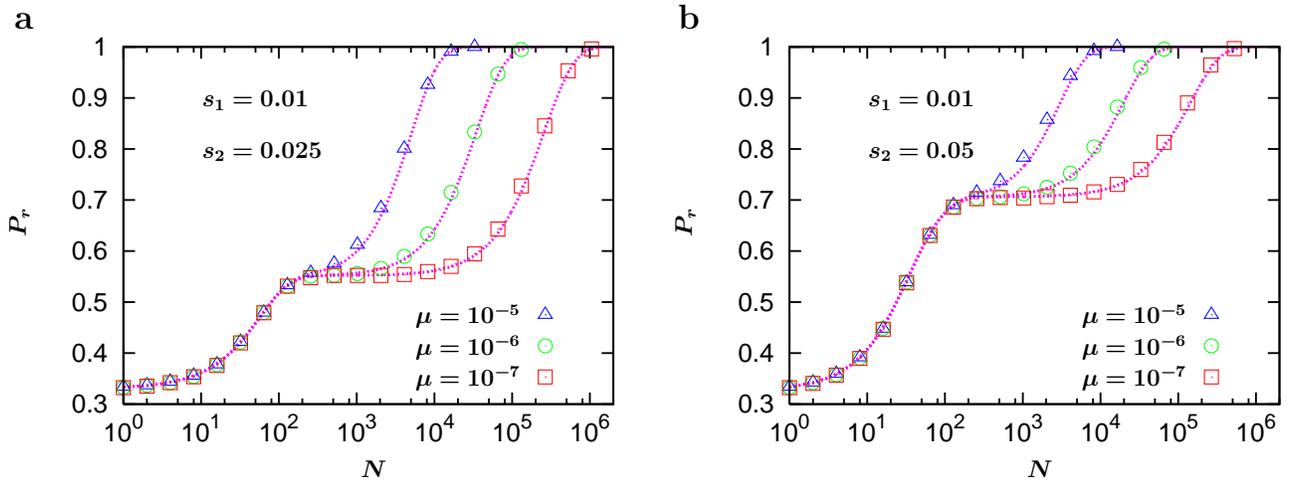}
\caption{\label{Prfig}Fixation probability $P_r$ of the allele $C$ for
the simplified three alleles system obtained using numerical
simulations 
for $\mu=10^{-5}$ (triangles), $10^{-6}$ (circles), and
$10^{-7}$ (squares). Dotted lines show the analytic prediction of 
equations (\ref{Pr_app},\ref{Q}).}
\end{figure}

To quantify the statement that large populations are more
likely to take the rugged path, we introduce the probability $P_r(N)$ that
the rugged path is taken by a population of size $N$. In simulations,
the probability $P_r$ was measured by 
counting the number of events in which $\{ 100 \}$ becomes
the most populated genotype for the first time. 
In Figure \ref{Prfig} these numerical
results are compared with the analytical expressions (discussed below) and
the two are seen to be in very good agreement.
We see that $P_r$ generally increases with $N$ 
(provided $\mu$ is not too small) thus supporting our main contention.  
Before presenting an
analytic calculation of $P_r$ covering the full range of population
sizes, we discuss the limiting cases of very small and very large
populations. 

\subsubsection*{SSWM regime: $N \mu \ll 1$}
 
When $N \mu \ll 1$, the path probability $P_r$ is equal to the 
probability that the first mutant that will be fixed in a population
initially monomorphic for the genotype $\{000\}$ is the local peak
genotype $\{100\}$. %Following \cite{Orr2002} this is given by 
%%PSC
That is,
\begin{equation}
\label{Pr_small}
P_r \vert_{N \mu \ll 1} = \frac{\pi(s_2)}{\pi(s_2) + 2 \pi(s_1)}.
\end{equation}
When selection is weak, in the sense of $N s_{1,2} \ll 1$, the
fixation probability is given by its neutral value $\pi = 1/N$
and we obtain $P_r = 1/3$. On the other hand, for strong selection
(and assuming that $s_{1,2} \ll 1$) we have 
\begin{equation}
\label{Pr_SSWM}
P_r^{SSWM} \approx \frac{s_2}{s_2 + 2s_1}
\end{equation}
independent of $N$,
which is equal to 0.54 and 0.56, respectively, in the two cases
displayed in Figure \ref{fit}. 

\subsubsection*{Deterministic quasispecies regime: $N \to \infty$}

For very large populations 
the local peak mutant is always present in
the population in considerable numbers and can therefore be expected
to dominate the population with a probability approaching unity
\citep{JK2007}.  
The quantitative analysis of this case is based on the deterministic
infinite population dynamics defined by equation
(\ref{quasispecies}). As the initial population is assumed to be fixed
at the genotype $\{ 000 \}$, for small mutation rates equation 
(\ref{quasispecies}) gives $f(\sigma,t=1) \sim \mu^{d(\sigma,000)}$ 
which is the same for all genotypes at constant Hamming
distance from $\{000\}$. 
For $t> 1$, the genotypic population can be 
determined by a simple construction described by \cite{Jain2005} and
\cite{Jain2007} 
in which the population of a genotype increases exponentially 
with its fitness, starting from $f(\sigma,1)$. 

For our fitness landscape, as the population frequencies 
$f(111,1) < f(100,1)$ but the 
fitness values $W(111) > 
W(100)$, it is possible that the genotype $\{111\}$ becomes the most populated
one before $\{100\}$ thus \textit{bypassing} the local maximum. 
%Since the frequency-increase of a genotype at local optimum by selection is much faster than that by mutation and the evolution of the whole populationis led by the frequency change of locally optimal genotypes(in the infinite population limit), neglecting the effect of mutations on frequency-increase for $t>1$can be a good approximation as illustrated by \cite{Jain2005} and \cite{Jain2007}.Thus, for $t>1$, we can assume that frequency of each genotypic population grows purely by selection, which results in$f(\sigma,t) \propto \mu^{d(\sigma,000)} W(\sigma)^t$ up to normalization.
The population at sequence $\{100\}$ overtakes
that of $\{000\}$ at time $t_1$ when $f(000,t_1)=f(100,t_1)$, which on
using $f(\sigma,t) \propto \mu^{d(\sigma,000)} W(\sigma)^t$ gives
$t_1=-\ln \mu/\ln W(100)$. Similarly the time $t_2$ at which the
population of the global maximum overtakes that of the initial
sequence is given by $t_2=-3 \ln \mu/\ln W(111)$. 
Thus the condition for bypassing corresponding to $t_2 < t_1$
reads 
\begin{equation} 
\label{bypass}
W^3(100) < W(111) \Leftrightarrow s_2 < s_1
\end{equation}
which is ruled out by construction. Thus bypassing cannot occur and  
we conclude that
\begin{equation}
\label{Pr_large}
\lim_{N \to \infty} P_r(N) = 1.
\end{equation}

\subsubsection*{Clonal interference regime}

The phenomenon of interest in this paper occurs in the intermediate 
range of population sizes where $P_r$ increases from its small
population value in equation (\ref{Pr_small}) to the large population
limit in 
equation (\ref{Pr_large}). This regime is more difficult to analyze because of the 
presence of multiple competing mutant clones in the population.
To find an analytic expression for $P_r(N)$ in this regime, 
we first reduce the three-locus
problem into a single locus with three alleles, say $A$ , $B$, and $C$
with respective fitness 1, $1+s_1$, and $1+s_2$. The two genotypes
$\{010\}$ and $\{001\}$ are lumped into a single allele $B$.
The mutation from $A$ to 
$B$ occurs with probability $2 \mu$ and that from $A$ to $C$
with $\mu$. No other mutation is possible, which ensures that
either $B$ or $C$ will be eventually fixed. It is clear that the fixation 
probability of allele $B$ approximates $1-P_r$.
%\subsection*{Calculation of $P_\text{\lowercase{$r$}}$}
%We consider the reduced model with three alleles $A$, $B$ and $C$ and
%fitness assignments 1, $1+s_1$ and $1+s_2$,
%respectively. Unidirectional mutations occur from $A$ to $B$ at rate
%$2\mu$ and from $A$ to $C$ at rate $\mu$.

We now present an approximate calculation of $P_r$ for the
three-allele model using ideas from clonal interference theory. 
At time zero the population is monomorphic
for allele $A$.
We would like to determine the probability that an allele $B$ which
originates at some time $t_0 > 0$ fixes at a later time $t_0+t_f$.  
It is plausible to assume that the fixation
and origination of a mutation are not correlated, and hence to 
treat the two processes separately.

Let us first consider the probability $p_1(t)$ for the allele $B$ to
originate at time $t$. An allele $B$ appears 
in the population at rate $2 N \mu$ and would, in the absence of 
other mutations, go to fixation with probability $\pi(s_1)$.
As is customary in the field \citep{MS1971,GL1998,DF2007},
we interpret the fixation probability $\pi$ as the probability for
the mutant population to survive genetic drift and, thus, to reach a
size large enough for the further time evolution to be essentially
deterministic. Mutations of type $B$ which reach this level are called 
contending
mutations \citep{Gerrish2001}, and they arise at rate 
$2 N \mu \pi(s_1)$. To obtain $p_1(t)$ this rate has to be
multiplied with the probability that the contending mutation in
question is the first to appear among all 
possible contenders for fixation. To estimate the probability that
no contending mutation (of any type) has appeared before time $t$,
we use a Poisson approximation in which the probability for the
non-occurrence of an event is the negative exponential of the expected
number of events. Since the expected number of contending mutations arising up to time 
$t$ is $N \mu (\pi(s_2)+ 2 \pi(s_1))t$, we conclude that
\begin{equation}
p_1(t) = 2 N \mu \pi(s_1) \exp(-N \mu (\pi(s_2)+ 2 \pi(s_1))t).
\end{equation}

We next determine the probability $p_2$ that the fixation of the contending
mutation of type $B$ is not impeded by the appearance of a contending
mutation of type $C$ at some time larger than $t_0$. Such a mutation can
only arise from the wildtype population $A$. According to our
assumptions, the evolution of the frequency $x$ of $B$ alleles 
after time $t_0$ follows the 
deterministic, logistic growth equation
\begin{equation}
\frac{d x}{dt} = s_1 x (1-x).
\end{equation}
The expected number of $C$ alleles that arise from the wildtype population
until the fixation time $t_f$ is therefore 
\begin{equation}
\int_0^{t_f} N \mu (1-x(t)) dt = \int_{x_0}^{1} \frac{N \mu }{x s_1} dx
= - \frac{N \mu}{s_1} \ln(x_0),
\end{equation}
where $x_0$ is the initial frequency of the contending $B$ allele.
As before, we use a Poisson approximation to determine the probability
that no contending mutation of type $C$ arises until the fixation of the
$B$ allele. Since the expected number of such contending mutations is
$-N \mu \ln (x_0) \pi(s_2) / s_1$, we have
\begin{equation}
p_2 = \exp[N \mu \ln(x_0) \pi(s_2) / s_1].
\end{equation}
% Note that as a first approximation we did not take into account the 
% increase of the mean fitness which decrease the fixation probability
% of $C$.
To obtain the total probability $1-P_r$ for the $B$ allele to fix we
multiply $p_1$ by $p_2$  and integrate over the initial time $t_0$, which gives
\begin{equation}
\label{Pr_deriv}
1 - P_r(N) = p_2 \int_0^\infty dt_0 \; p_1(t_0) = \frac{2 \pi(s_1)}{\pi(s_2)+ 2 \pi(s_1) }
e^{N \mu \ln(x_0) \pi(s_2) / s_1}.
\end{equation}

To complete the analysis we have to determine $x_0$. A naive argument
would suggest that $x_0 = 1/N$ because the contending mutation should start from
a single mutant. However, this does not include the fact that the
fixation process \textit{conditioned on survival} 
is faster than the logistic growth with $x_0=1/N$
 \citep{HP2005,DF2007}. 
A simple approximate way to take into account this effect is to 
let the contending mutant clone start at frequency $x_0 = 1/(s_1 N)
\gg 1/N$ \citep{MS1971}. Inserting this into equation (\ref{Pr_deriv}) we
obtain the final  result given by  
%In \textbf{Methods} we present an approximate calculation of
%$P_r$ for the three-allele model using ideas from clonal interference
%theory. The result is
\begin{equation}
P_r (N)= 1 - \frac{2 \pi(s_1)}{\pi(s_2)+ 2 \pi(s_1) }
\exp(-Q(N)),
\label{Pr_app}
\end{equation}
where 
\begin{equation}
\label{Q}
Q(N) = N \mu \ln (N s_1) \pi(s_2) / s_1.
\end{equation}
Figure \ref{Prfig} shows that  equation
(\ref{Pr_app}) agrees well with simulations of the three-allele model
described at the beginning of this section. 
In the strong selection limit, using $\pi(s) \approx 2s$ the above
expression for $P_r$ can be simplified to give $P_r(N) \approx 1-
(1-P_r^{SSWM}) e^{-Q(N)}$ with $Q(N) \approx 2 N \mu \ln (N s_1)
s_2/s_1$. Thus the path probability is close to the  
SSWM value when $Q(N) \ll 1$ and approaches unity for
large $Q(N)$. The increase of $P_r$ beyond the SSWM value which
ultimately gives rise to the reversal in the ordering of fitness with
increasing population size in Figure \ref{fit},  takes place when
$Q(N)$ is of order unity,
\begin{equation}
\label{CI}
N \mu \ln(N s) \sim {\cal{O}}(1)
\end{equation}
where it is assumed that both selection coefficients have a similar
scale $s$. 
%Thus for population sizes of order $c/\mu \ln (cs/\mu)$, 
%Equation (\ref{Pr_app}) interpolates smoothly between the limits givenby equation (\ref{Pr_small}) and equation (\ref{Pr_large}), and agrees wellwith simulations of the three-allele model (Figure \ref{Prfig}). For small $\mu$ a plateau corresponding to theSSWM value in equation (\ref{Pr_SSWM}) appears. 

%The increase of $P_r$ beyond its SSWM value in equation (\ref{Pr_small}), which ultimatelygives rise to the reversal in the ordering of fitness with increasingpopulation size in Figure \ref{fit}, takes place when $Q(N) = {\cal{O}}(1)$, that is, when\begin{equation}\label{CI}N \mu \ln(N s) \sim 1,\end{equation}where it is assumed that both selection coefficients have a similarscale $s$. The region in the $(N,\mu)$-plane where a transient advantage of small populationscan be expected is therefore limited by lines of constant $Q$. This is illustratedin Figure \ref{Qplot}, where the limiting values have been set (somewhat arbitrarily) to $Q = 0.1$ and $Q = 10$. For $\mu = 10^{-5}$ one reads off from the figure that small populationsare at an advantage in the range $1000 \leq N \leq 20000$, which is in reasonable agreement withthe simulations shown in Figure \ref{fit}. Note that, since $\pi(s_2)/s_1 \approx 2 s_2/s_1 \approx 5$ in bothpanels of the figure, the predicted range is the same in the two cases.  

It is straightforward to generalize the above derivation of $P_r$ to
an $L$-locus 
system where $L-1$ initial mutations confer a selective advantage
$s_1$ and one mutation confers a higher advantage $s_2 > s_1$. This
merely increases the mutation rate from allele $A$ to allele $B$ to 
$(L-1)\mu$ and leads to the expression 
\begin{equation} 
P_r(N) = 1 - \frac{(L-1) \pi(s_1)}{\pi(s_2)+ (L-1) \pi(s_1) }
\exp(-N \mu \ln(Ns_1) \pi(s_2) / s_1).
\end{equation}

\subsection*{House-of-cards models}

At this point the question naturally arises as to how 
generic our results are with respect to the number of loci and the
structure of the fitness landscape. 
To address this question, 
we simulated populations evolving in 
two ensembles of random fitness landscapes, the unconstrained
and constrained house-of-cards models. 
In these models the fitness values of genotypes differing by single
mutational steps are assumed to be uncorrelated. While this is not
likely to be the case in real fitness landscapes \citep{Miller2011},
the house-of-cards models constitute the conceptually simplest realization of a generic, rugged fitness
landscape which is essentially parameter-free, apart from the overall
fitness scale $S$ in equation (\ref{Eq:fit_HOC}). 

Before discussing the dynamics of adaptation in random landscapes, we may ask  
how typical the three-locus landscape equation (\ref{model}) itself is within
the constrained ensemble with $L=3$. The main topographic features of the landscape equation (\ref{model})
are (i) the existence of a single local maximum, in addition to the global maximum, and 
(ii) the existence of 2 rugged and 4 smooth paths from the global minimum $\{000\}$ to 
the global maximum $\{111\}$. The enumeration of all $6! = 720$ possibilities of ordering the
fitness values of the 6 genotypes intermediate between the global minimum and the global maximum 
shows that these two features are shared by a fraction of $11/120 \approx 0.0917$ of all landscapes in
the constrained ensemble. 
However, the evolutionary dynamics of populations is determined not only
by the order statistics of the fitness landscape, but depends also on the
magnitude of the selection coefficients assigned to different adaptive pathways. 
These effects can only be addressed by explicit simulations, which we discuss next.

%\subsection*{House-of-cards model with three loci}

\begin{figure}
\centerline{\includegraphics[width=1.0\textwidth]{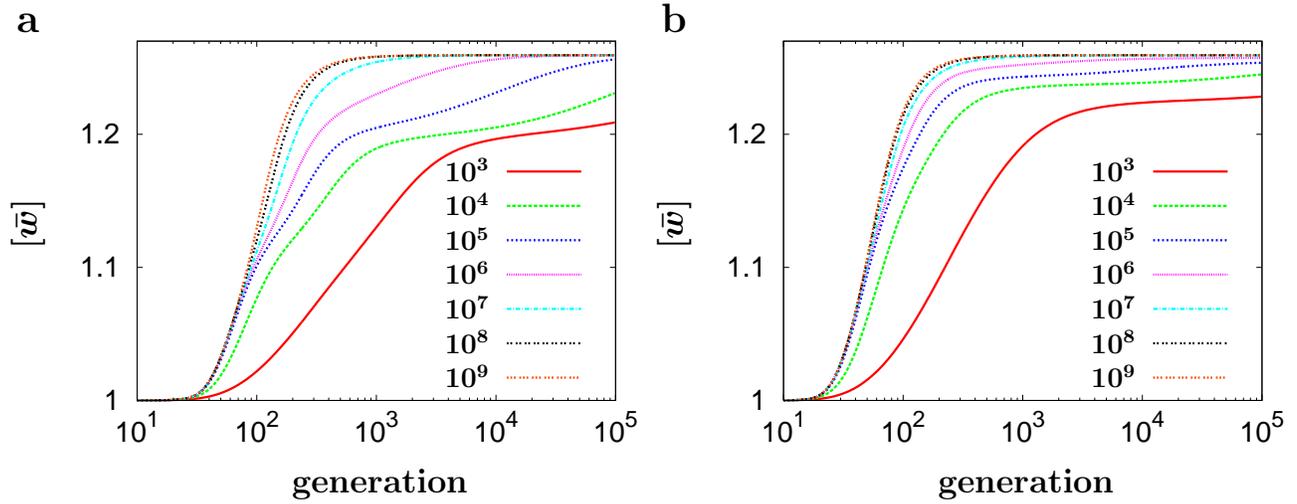}}
\caption{\label{Fig:meanfit} Mean fitness evolution in random
  three-locus fitness landscapes.
Fitness has been averaged over $10^4$ realizations of the landscape and 
128 population histories in each realization, with parameters $\mu =
10^{-5}$ and $S = 0.1$, and population sizes as indicated in the figures. 
Both for (a) the constrained ensemble and (b) the unconstrained ensemble
the mean fitness increases monotonically with population size for all times.} 
\end{figure}

In Figure~\ref{Fig:meanfit} we show the time evolution of the mean fitness, averaged over 
many realizations of the landscape, for both ensembles of random three-locus landscapes; as before,
all individuals are initially placed at the least fit genotype. In contrast to Figure~\ref{fit}, the 
mean fitness is now a monotonic function of population size at all times: When averaged over 
the landscape ensemble, there is no advantage of small populations. This however does not 
preclude the existence of such an advantage for individual landscapes. To test for this 
possibility, we introduce the probability $P(t,N,N')$ of a small population
advantage, defined as the probability that the mean
fitness of a population of size $N$ at generation $t$ is larger
than that of a population of size $N'$ at the same generation, where $N < N'$.

To determine $P(t,N,N')$ numerically, we first calculated 
the mean fitness for population size $N$ by averaging over 
128 independent evolutionary histories on a single landscape, 
and then compared it to that for a different population size $N'$ 
on the same landscape. Finally the outcome of the comparison was 
averaged over $10^4$ samples of the random landscape ensemble.  
In order to avoid spurious contributions from cases where the fitness is 
in fact independent of population size and an apparent advantage of small populations
arises due to fluctuations, we count only instances 
in which the inequality  $(1-\alpha) w(N)>  w(N')$ is satisfied,
where $w(N)$ stands for the calculated mean fitness for population size $N$. 
We choose $\alpha = 10^{-3}$, which is large enough to remove the error
due to fluctuations. Of course, this may also screen out landscapes
with very small advantage, but we do not think that this effect is 
substantial, since a mean fitness advantage of less than 0.1\%
is negligible compared to the scale $S = 0.1$ of typical selection coefficients.  
% If the mean fitness advantage is less than 0.1\%,
% it is not a serious error to say that there is no advantage.

\begin{figure}
\centerline{\includegraphics[width=1.0\textwidth]{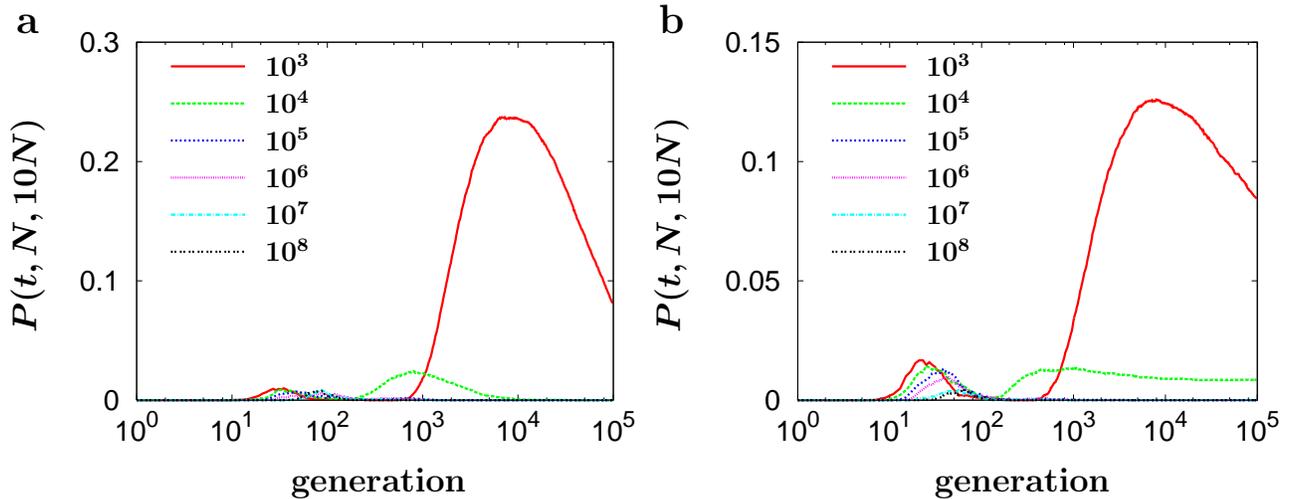}}
\caption{\label{Fig:mean1e3} Probability of small population advantage in random 
three-locus fitness landscapes. Plots show
$P(t,N,N')$ as a function of $t$, with $N'=10N$ and $N$ as indicated
in the figures, $\mu=10^{-5}$, and $S = 0.1$. Part (a) shows the constrained ensemble, part (b) the 
unconstrained ensembles. 
The constrained ensemble with the maximum
fitness at the antipodal point is more likely to allow small
populations to have larger mean fitness.
}
\end{figure}

%PSC
The resulting probabilities for the two ensembles
are summarized in Figure~\ref{Fig:mean1e3}.
Referring first to the constrained ensemble, which mimics most closely
the overall topography of the landscape equation (\ref{model}),
we see that a  conspicuous probability for an advantage of small populations
begins to appear around $10^3$ generations and is most pronounced for 
population size $N \approx 10^3$. Note that in Fig.~\ref{fit} {\bf a} where the 
selection coefficient is of the order of 0.1 as in Fig.~\ref{Fig:mean1e3}, 
the fitness advantage of a small population becomes conspicuous when the 
population with size $10^3$ is compared to that with $10^4$ in the
time window between $10^3$ and $10^5$. That is, the quantitative as well as 
the qualitative behavior of the house-of-cards model in the constrained ensemble
is similar to the three-locus model in the previous section.
%in close agreement with the behavior in Figure~\ref{fit}. 
Thus, we may conclude
that the landscape equation (\ref{model}) is quite generic within the constrained ensemble,
in the sense that between 20\% and 30\% of all landscapes show similar
dynamical features. In the unconstrained ensemble the probabilities
are reduced by about a factor of 2, but the overall behavior is the same.

\subsection*{More than three loci}

Within the house-of-cards models, it is straightforward to investigate
the effect of varying the number of loci $L$.  In the simulations
reported in this subsection 
we allow for at most one mutation per individual and generation, and specify
the genome-wide mutation probability $U$ rather than the mutation 
probability $\mu$ per locus. The general relation 
between $U$ and $\mu$ is \citep{Park2008}
$
U = 1 - (1 - \mu)^L \approx \mu L
$ when $\mu L \ll 1$.
With increasing $L$
the difference between the constrained and unconstrained ensembles
becomes less important, because populations typically do not reach
the global maximum within the simulation time, and hence its precise 
location is irrelevant. In the following we therefore restrict the 
discussion to the standard (unconstrained) model.

\begin{figure}
\centerline{\includegraphics[width=1.0\textwidth]{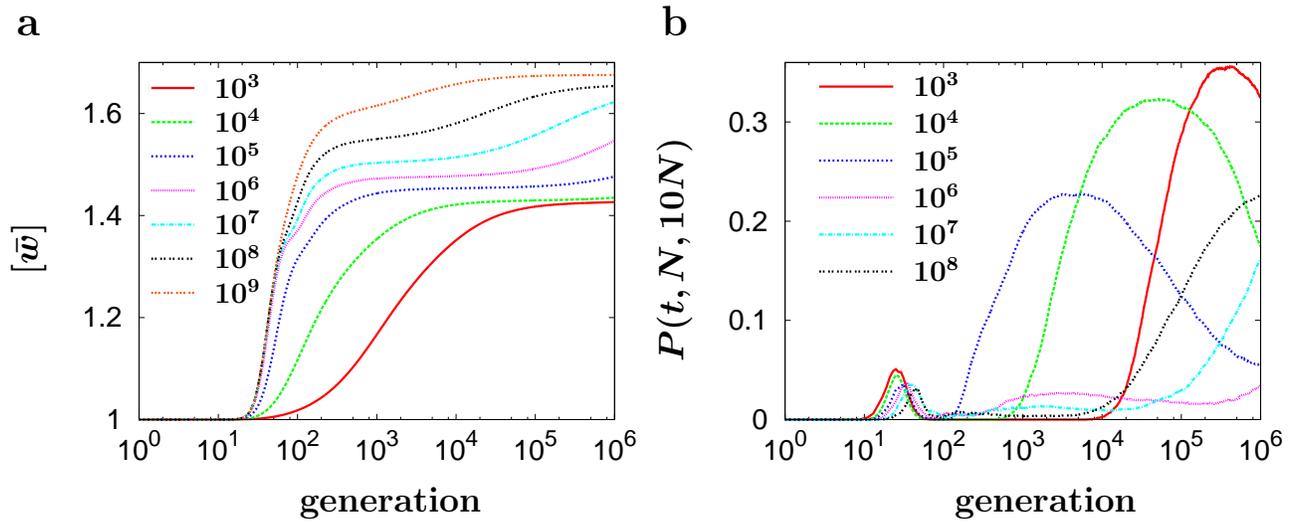}}
\caption{\label{Fig:20_1e3} (a) Mean fitness evolution  and (b) probability
of small population advantage for the unconstrained
ensemble with $L=20$ loci. Parameters are $U=10^{-5}$ and $S=0.1$, and
population sizes as indicated in the figures. 
}
\end{figure}

As an illustration, we present simulation results for $L=20$ in
Figure~\ref{Fig:20_1e3}. In this case the expected fitness value of the
global maximum is $\approx 2.44$ which is far larger than the mean
fitness of the population with size $10^9$ within the
observation time. This means that most of evolutions up to size $10^9$ and generation $10^6$
have not arrived at the global maximum, but rather explore the surroundings 
of the low fitness starting genotype. As in the three-locus case, the mean fitness
averaged over all landscapes is monotonic in the population size 
(Figure~\ref{Fig:20_1e3}a), and the plot of $P(t,N,N')$ in  
Figure~\ref{Fig:20_1e3}b displays a similar, though more pronounced advantage of 
small populations for $N = 10^3-10^5$. However, for $N \geq 10^7$ a new peak
is seen to appear in $P(t,N,N')$ at later time, which is not present for
$L=3$. 

We may interpret this behavior in terms of the different evolutionary regimes
described by \cite{JK2007}. The disappearance of the first peak in $P(t,N,N')$
marks the point where the mutation supply rate $NU$ is sufficiently large for
the population to easily escape from local fitness maxima by the creation
of double mutants. Such a population will however still have difficulties to cross
wider fitness valleys, and hence
it will tend to get trapped at local maxima which are separated  
by three or more mutational steps. The mechanism for an advantage of small
populations that was found in the three-locus model 
% (see equation (\ref{model})) 
thus reemerges on a larger scale, giving rise to secondary peaks in $P(t,N,N')$.

%==========================================================================
\section*{Discussion}

Understanding 
the effect of population size on the rate of adaptation is a central
problem in evolutionary theory, which continues to attract
considerable attention \citep{Weinreich2005b,Gokhale2009,Weissman2009,Lynch2010}. Motivated by the experiments
of \cite{Rozen2008}, the present work has addressed a specific aspect
of this general problem. 
In the experiments, several populations of {\it
  E. coli} consisting of either $5 \times 10^5$ or $2.5 \times 10^7$ 
individuals were evolved in a complex nutrient medium which can be modeled by a complex fitness
landscape. The fitness measurements after $500$ generations showed
that small populations can achieve higher fitness than large
populations.  

A classic scenario in which 
a small population can acquire an evolutionary advantage
because of genetic drift has been put forward in the framework
of Wright's shifting balance theory \citep{Wright1931},
%, and it may be tempting to employ this theory
referred to as SBT in the following discussion.  
%to explain the experimental observation of \cite{Rozen2008}. 
%However, for a number of reasons the SBT as such cannot account for the experimental results.
Apart from the intrinsic shortcomings of the SBT \citep{Coyne1997}, however,
there are several reasons why it cannot be directly applied to the experimental situation of \cite{Rozen2008}.
First of all, the population in the experiments is not structured, 
and it is therefore not possible for different demes to occupy separate fitness peaks
as assumed in phase II of the SBT. Second, the number of generations in the experiment
is too small for the entire population to cross a fitness valley either
by the fixation of deleterious mutations (phase I of the SBT) or by  
the simultaneous fixation of individually deleterious but jointly beneficial mutations
\citep{Weinreich2005b}. 
% However as equation (\ref{Tesc}) shows, valley crossing by simultaneous 
% fixation of mutations takes more time for a small population than for a large one. 
% Moreover, the time scale of valley crossing is $O(\mu^{-2})$ where
% $\mu$ is the mutation probability~\citep{Weinreich2005b}, so it does not 
% seem possible for the population to cross the valley within 100 generations
% which is the time scale of the experiment \citep{Rozen2008}.
% as seen in the experiment.
Hence for a proper explanation of the experimental observations 
it cannot be assumed that the population resides 
at a local fitness optimum from the beginning of the process. 
Rather, the evolutionary trajectories begin in a fitness valley, and 
the dynamics is determined by the competition between 
different fitness peaks that are available to the population \citep{Rozen2008}. 

In the preceding
sections we have demonstrated that under these conditions, 
small populations may indeed reach higher 
fitness levels than large ones  because they are more likely to evade
trapping at local fitness maxima. 
%We first considered a simple case of a diallelic three-locus model to develop a quantitativeanalytic theory which, in particular, allows us to predictthe range of population sizes and mutation rates in which themechanism operates. We also studied ...
Our detailed study of a three-locus model where a single local
fitness peak competes with the  global maximum has shown that 
%and  identified parameter regimes in which the fitness of a small
%population can overtake that of a larger one. 
the dynamical behavior
of the population fitness is not determined by the time scale to
acquire beneficial mutation(s) alone and 
%presented a detailed analysis based on 
depends on the path probability $P_r(N)$ also. This probability 
increases from a constant value $1/3$ to $P_r^{SSWM}$
and finally approaches the deterministic value unity as the population
size $N$ increases. For the parameter regimes in which $P_r$ is
constant in $N$, as the waiting times $T_{\mathrm{rugged}}$ and
$T_{\mathrm{smooth}}$ both decrease with $N$, larger populations are at
an advantage. 

However when the probability to take a rugged path
increases with $N$, a larger population may get trapped at the local
fitness maximum thus acquiring lower fitness than a small
population. For the
parameters in 
Figure~\ref{fit}, the path probability exceeds the SSWM value $0.55$
for $N > 500$. From equation (\ref{Pr_app}), it is approximately
$0.6$ for $N=10^3$ but close to unity for 
$N \geq 10^4$ and therefore a population of size $N \sim 10^3$ is able to
acquire a higher fitness  than larger populations.

A key result of our analysis is that the
regime of population sizes in which this mechanism operates coincides
with the onset of clonal interference, which occurs precisely when 
the criterion in equation (\ref{CI}) is satisfied
\citep{GL1998,deVisser1999,W2004,Park2007}. 
In the context of the three-allele model discussed previously, 
clonal interference implies that a high fitness
clone (allele C) may arise while the low fitness mutant (allele B) is still on the way to fixation, thus
enhancing the probability for C to fix and increasing the probability for the 
population to evolve along a rugged path.
%For $N s \gg 1$, path probability changes in the clonal interference
%regime defined by equation (\ref{CI}). 
 From the criterion in
equation (\ref{CI}), we 
can determine the mutation probability 
for the bacterial populations used in the experiment of
\cite{Rozen2008}. 
Since the small population advantage is observed around $N= 5 \times 10^5$ and
the characteristic size of selection coefficients derived from the
fitness trajectories in the experiment is $\simeq 0.1$, 
the estimated mutation probability is
\begin{equation}
\mu \approx \frac{1}{N \ln(Ns)} \simeq 10^{-6},
\label{mu}
\end{equation}
which should be interpreted as a \textit{beneficial} mutation rate.
This estimate is consistent with values for  
the beneficial mutation rate in \textit{E.coli}
obtained by other experimental approaches, 
which range from $10^{-4}$ to $10^{-7}$
\citep{Hegreness2006,PFMG2007}. 

Our theoretical analysis also shows
that the advantage of small populations over large ones is
transient and at sufficiently large times, the fitness of the large
population exceeds that of the small 
population. This reversal occurs  when the large population escapes the
fitness valley at time $T_\mathrm{esc} \sim (N \mu^2)^{-1}$ (see
Eq.~\ref{Tesc}) and should be testable over an
experimentally accessible time scale of $10^3$ generations if the
population size exceeds $(10^{3} \mu^2)^{-1} \sim 10^9$ where we have
used (\ref{mu}). In the
experiments of \cite{Rozen2008}, the large populations had a size of $N \sim
10^7$ which is two orders of magnitude below our prediction and hence
the reversal in the ordering of fitness was not observed. It would be
interesting to test this effect in experiments using larger populations.

In their work, \cite{Rozen2008} attributed the fitness
advantage seen for small populations  to the  
heterogeneity in evolutionary  trajectories. This qualitative
description can be made precise by considering 
%The behavior of $P_r(N)$ can be rephrased in terms of the
predictability of the first adaptive step along the evolutionary
trajectory. Following \cite{Orr2005b} and \cite{Roy2009} we define the 
predictability $P_2$ to be the probability that two replicate 
populations follow the same evolutionary trajectory, which equals 
the sum of the squares of the
probability of these trajectories. Specializing to the 
first adaptive step in our three-locus system, the possible outcomes 
$\{100\}$, $\{010\}$ and $\{001\}$ occur with probability $P_r$, 
$(1-P_r)/2$ and $(1-P_r)/2$, respectively so that $P_2=P_r^2 +
\frac{1}{2} (1 - P_r)^2$
%. The expression for the 
%predictability $P_2$ thus reads
%\begin{equation}
%\label{predict}
%P_2 = P_r^2 + \frac{1}{2} (1 - P_r)^2,
%\end{equation}
which increases from $P_2 = 1/3$ to $1$ with increasing population
size, similar to the behavior of $P_r$ itself. For the parameter range
$N \mu \ll 1, N s \gg 1$ 
in which $P_r$ is independent of $N$, the predictability $P_2 < 1$ is
also a constant and as discussed above, we do not
expect a fitness advantage for such populations. Thus
although small populations will exhibit heterogeneous evolutionary
trajectories in general, this does not necessarily imply a fitness advantage.

The robustness of our results with respect to modifications of the
fitness landscape equation (\ref{model}) was addressed within the
framework of two variants of the house-of-cards model. 
From this study we concluded
%From the above study of two variants of the house-of-cards model, we conclude 
that an advantage of small populations is generally not present when
the mean fitness is averaged over all possible realizations of the landscape;
however, in the appropriate range of time scales and population sizes
identified 
for the landscape equation (\ref{model}), a non-negligible fraction of
all landscapes 
do display such an advantage. This agrees with the results obtained by 
\cite{Handel2009} for a class of models with tunable ruggedness, 
and is also consistent with the analysis of the mean
fitness evolution in the infinite sites limit of the house-of-cards
model by \cite{Park2008}. 

Returning to the evolution experiments that initially motivated this work, 
we expect the effects described here to be observable whenever the underlying
fitness landscape displays multiple local maxima. Among the small number of 
available empirical fitness landscapes, local maxima have been found in some cases
\citep{dVPK2009,Lozovsky2009} but not in others \citep{Weinreich2006}. 
Further work on the structure of empirical fitness landscapes is
therefore needed in order to assess the generality of the proposed mechanism
for an evolutionary advantage of small populations.

%%%%%%%%%%%%%%%%%%

%%%%%%%%%%%%%%%%%%%%%%%%%%%%%%%%
%\section*{Authors contributions}
%
%K.J. and J.K. planned the study. K.J. and S.C.P. carried out numerical
%and analytic calculations. K.J., J.K. and S.C.P. wrote the paper.

%%%%%%%%%%%%%%%%%%%%%%%%%%%
\subsection*{Acknowledgements}
This work was supported by DFG within SFB 680 \textit{Molecular Basis
of Evolutionary Innovations}. We thank Arjan de Visser, Siegfried Roth
and Sijmen Schoustra for helpful discussions and comments, and two anonymous reviewers for their
suggestions on the manuscript. 
% and for pointing out \citep{Handel2009} and \citep{Carneiro2009}. 
K.J. and J.K. acknowledge the hospitality of 
KITP, Santa Barbara, and support under NSF grant PHY05-51164
during the initial stages of this work.

%\bibliographystyle{evolution} 
%\bibliography{me}

%%%%%%%%%%%

\subsection*{Figures}

%{\bf Here are some suggestions for revising the figures: (i) drop 1a
%(of current version) as 1b parameters have been used in Fig 2. (ii)
%display 2a as inset of 2b (right bottom corner) (iii) drop either Fig
%3a or 3b (iv) mv results on unconstrained model to online appendix
%(v) display 4a as inset of 5 (vi) display 6a as inset of 6b. In some
%of these cases, the labeling of parameter N can be shifted to caption.}

\end{document}